\documentclass[12pt]{article}
\usepackage{amsmath}
\usepackage{amssymb}
\usepackage{amscd}
\usepackage[dvips]{graphicx}
 \linethickness{0.4pt}
\begin{document}
~~
\bigskip
\bigskip
\bigskip
\bigskip
\bigskip
\begin{center}

\section*{{%\large\bf
TOWARDS QUANTUM NONCOMMUTATIVE $\kappa$-DEFORMED FIELD
THEORY}\footnote{Supported by KBN grant 1P03B01828.}}
\end{center}
\bigskip
\bigskip
\bigskip
\begin{center}
{{\large\bf ${\rm Marcin\;Daszkiewicz}$, ${\rm Jerzy\; Lukierski}$,
${\rm Mariusz\;Woronowicz}$}}
\end{center}
\begin{center}
{%\large
{Institute of Theoretical Physics\\ University of Wroc{\l}aw pl.
Maxa Borna 9, 50-206 Wroc{\l}aw, Poland\\ e-mail:
marcin,lukier,woronow@ift.uni.wroc.pl}}
\end{center}
\bigskip
\bigskip
\bigskip
\bigskip
\bigskip
\bigskip

\begin{abstract}
We introduce new $\kappa$-star product describing the multiplication
of quantized $\kappa$-deformed free fields. The $\kappa$-deformation
of local free quantum fields originates from two sources:
noncommutativity of space-time and the $\kappa$-deformation of field
oscillators algebra - we relate these two deformations. We
demonstrate that for suitable choice of $\kappa$-deformed field
oscillators algebra the $\kappa$-deformed version of microcausality
condition is satisfied, and it leads to the deformation of the
Pauli-Jordan commutation function defined by the $\kappa$-deformed
mass shell. We show by constructing the $\kappa$-deformed Fock space
that the use of $\kappa$-deformed oscillator algebra permits to
preserve the bosonic statistics of n-particle states. The proposed
star product is extended to the product of $n$ fields, which for
$n=4$ defines the interaction vertex in perturbative description of
noncommutative quantum $\lambda\varphi^4$ field theory. It appears
that the classical fourmomentum conservation law is satisfied at the
interaction vertices.
\end{abstract}
\bigskip
\bigskip
\bigskip
\bigskip

\eject

\section{Introduction.}

Noncommutative space-time geometry was firstly physically motivated
by combining together the postulates of general relativity and
quantum mechanics \cite{lit1}, \cite{lit2}; alternatively the
noncommutative coordinates were derived in brane-world scenario as
describing the space-time manifolds attached to the ends of open
strings \cite{lit3}, \cite{lit4}. As next step it is important to
construct in noncommutative space-times the classical and quantum
dynamical models, in particular, the noncommutative counterparts of
quantum local fields and their perturbative expansions.

In noncommutative deformed QFT there are two possible sources of
modifications of the standard formulae for quantum fields:

i) We replace the classical Minkowski space-time describing in
standard theory the local field arguments by the noncommutative
space-time coordinates (see e.g. \cite{lit5}, \cite{lit6})
\begin{equation}
[\;{\hat x}_{\mu},{\hat x}_{\nu}\;] =
\frac{i}{\kappa^2}\theta_{\mu\nu}(\kappa x_{\mu}) =
\frac{i}{\kappa^2}\theta_{\mu\nu}^{(0)} +
\frac{i}{\kappa}\theta_{\mu\nu}^{(1)\rho}x_{\rho} + \cdots\;,
\label{noncomm}
\end{equation}

ii) Second change comes from the introduction of modified field
quantization rules which replace the known infinite set of canonical
creation and annihilation operators. In such a way we modify
standard boson/fermion oscillators into the $\kappa$-deformed ones.

Recently, there was studied the structure of free quantized
noncommutative fields for the "canonical" deformation
$\theta_{\mu\nu} \equiv \theta_{\mu\nu}^{(0)}$. Important step in
understanding of such a theory was the discovery of Hopf-algebraic
twisted quantum symmetry leaving invariant the noncommutativity
relation (\ref{noncomm}) with $\theta_{\mu\nu} =
\theta_{\mu\nu}^{(0)}$ \cite{lit7}-\cite{lit8a}. It was shown in
such a case (see \cite{lit10}-\cite{lit11a}) that the
$\theta_{\mu\nu}^{(0)}$-deformation of quantum free field implies
the modification of the field oscillators algebra. It appears that
perturbative Gell-Mann-Low expansion for interacting
$\theta_{\mu\nu}$-deformed quantum $\lambda\phi^4$ theory can be
identified with the analogous undeformed expansion, obtained by
putting $\theta_{\mu\nu} = 0$ \cite{lit10}, \cite{lit11},
\cite{lit11a}.

In this paper we shall consider QFT on noncommutative Lie-algebraic
space-time (see (\ref{noncomm}) with $\theta_{\mu\nu}^{(1)\rho} \ne
0$), in particular on $\kappa$-deformed Minkowski space
\cite{lit13}-\cite{lit15}
\begin{equation}
[\;{\hat x}_{0},{\hat x}_{i}\;] = \frac{i}{\kappa}{\hat
x}_{i}\;\;,\;\; [\;{\hat x}_{i},{\hat x}_{j}\;] = 0\;.
\label{minkowski}
\end{equation}
Our main aim is to formulate the $\kappa$-deformation of the theory
of free quantum fields in a way implying the $\kappa$-deformed
version of microcausality condition and to provide a ground for the
application to interacting case. The perturbative formulation of
interacting $\kappa$-deformed QFT as well as deformed Feynman
diagrams has been considered previously (see e.g.
\cite{lit16}-\cite{lit17b}), however without considering the
$\kappa$-deformation of the field oscillators. In particular, two
problems important for the construction of physically plausible set
of $\kappa$-deformed Feynman diagrams were not satisfactorily
solved:

$\alpha$) How to introduce the $\kappa$-star multiplication
$\star_{\kappa}$ of the free quantum fields which leads to
$\kappa$-version of the microcausality relation\footnote{The
standard version of causality condition in $\kappa$-deformed theory
has to be modified, because the notion of standard light cone is
modified under $\kappa$-deformation.}. For $\kappa$-deformed free
fields which can be decomposed into positive and negative frequency
parts $\varphi (x) = \varphi^{(+)} (x) + \varphi^{(-)} (x)$ we
should get $([\;A,B\;]_{\star_{\kappa}} := A\star_{\kappa} B -
B\star_{\kappa} A)$
\begin{equation}
[\;\varphi^{(\pm)} (x),\varphi^{(\pm)} (y)\;]_{\star_{\kappa}} =
0\;, \label{micro}
\end{equation}

$\beta$) How to define the local $\kappa$-deformed field vertices
which provide the classical Abelian fourmomentum conservation law
for ingoing and outgoing particles. In such a way physically
difficult to accept non-Abelian quantum addition law of the
fourmomenta will not appear in the description of the particle
scatterings.

An important step which permitted us to solve the problems $\alpha$)
and $\beta$) was the construction of $\kappa$-deformed algebra of
creation and annihilation operators proposed in \cite{lit18}. It is
easy to see that the $\kappa$-deformed addition law for the
fourmomenta due to the nonsymmetric fourmomentum coproduct requires
the modification of standard oscillators algebra. It follows (see
also \cite{lit18}) that after the exchange of two $\kappa$-deformed
particles their threemomentum dependence is changed by the
multiplicative factors depending on the energy of the other
particle. We shall show that the $\kappa$-deformed algebra of
oscillators for relativistic free fields in comparision with the
case considered in detail in \cite{lit18} should be suitably
generalized by the choice of numerical normalization factors. We
shall also argue that if we use in particular way the
$\kappa$-deformed oscillators algebra one can create from the vacuum
the multiparticle states with classical fourmomentum addition law,
and obtain the n-particle states obeying standard bosonic symmetry
properties.

The plan of our paper is the following. In Sect. 2 we shall describe
the known Hopf-algebraic framework of $\kappa$-deformed relativistic
symmetries which shall be used in our considerations. For the
definition of $\kappa$-deformed noncommutative free field we shall
employ the "symmetric" choice of the noncommutative Fourier
transform (\cite{lit21ago}, \cite{ago}). In Sect. 3 we shall
introduce new $\kappa$-star product describing the multiplication of
free quantized $\kappa$-deformed local fields. We stress here that
our new $\star$-multiplication is a novelty: besides representing
the multiplication of noncommutative $\kappa$-Minkowski coordinates
it introduces additional $\kappa$-dependent change of the mass-shell
conditions in the multiplied fields. We shall show that with such a
new multiplication $\star_{\kappa}$ the commutator is a c-number and
one obtains for free quantum $\kappa$-deformed K-G fields the
following commutation relations
\begin{equation}
[\;\varphi (x),\varphi (y)\;]_{\star_{\kappa}} =
\frac{1}{i}\Delta_\kappa (x-y;M^2)\;, \label{micro1}
\end{equation}
where the $\kappa$-deformed Pauli-Jordan function $\Delta_\kappa
(x;m^2)$ is described by the $\kappa$-deformed mass-shell condition
\begin{equation}
\Delta_\kappa (x;M^2) = \frac{i}{(2\pi)^3} \int d^4p \epsilon(p_0)
\delta \left(\left (2\kappa\sinh
\left({p_0}/{2\kappa}\right)\right)^2 - \vec{p}^{\ 2} - M^2 \right)
{\rm e}^{i{p}_\mu{x}^{\mu}}\;. \label{pauli}
\end{equation}
The relation (\ref{micro1}) completes the relations (\ref{micro})
and provide the explicit example of quantum field operator
satisfying the $\kappa$-causality condition. Leaving more detailed
discussion of the $\kappa$-causality to our future investigations we
recall that the commutator (\ref{micro1}) was firstly proposed in
\cite{lit19}. In Sect. 4 we shall introduce new multiplication
operation in the algebra of field oscillators and shall describe the
full algebra of $\kappa$-deformed creation and annihilation
operators. In Sect. 5 we shall introduce the corresponding
$\kappa$-deformed Fock space with n-particle particle states
carrying fourmomenta which add accordingly to classical Abelian
addition law. Further, in Sect. 6 we consider the $D=4$ local
$\kappa$-deformed vertex
\begin{equation}
\lambda \varphi^4(x) \rightarrow
\lambda\varphi(x)\star_{\kappa}\varphi(x)\star_{\kappa}\varphi(x)\star_{\kappa}\varphi(x)\;,
\label{vertex}
\end{equation}
which in momentum space is proportional to the Dirac delta
describing classical conservation law of fourmomenta. Finally, in
Sect. 7 we provide final remarks.

\section{$\kappa$-deformed Poincare symmetries and noncommutative Fourier transforms.}

The Hopf-algebraic description of $\kappa$-deformed Poincare
symmetries with mass-like fundamental deformation parameter $\kappa$
was introduced in \cite{lit19}, \cite{lit20}, where the standard
basis with modified boost sector of Lorentz algebra was proposed. By
nonlinear transformation of boost generators one arrives at the
bicrossproduct basis \cite{lit14} with classical Lorentz algebra
sector. Here we shall consider a modified bicrossproduct basis,
which was discussed earlier in \cite{lit21ago}, \cite{ago},
\cite{ArMar} with standard $\kappa$-deformed mass-shell condition
(see \cite{lit19}) and classical formulae for the coinverses.
Introducing the Poincare algebra generators $M_{\mu\nu} = (M_i =
\frac{1}{2}\epsilon_{ijk}M_{jk}, N_i = M_{i0})$ and $P_{\mu} =
(P_i,P_0)$ we get the following Hopf algebra relations

a) algebraic sector
\begin{eqnarray}
&&[\;M_{\mu\nu},M_{\lambda\sigma}\;] = i\left(
\eta_{\mu\sigma}M_{\nu\lambda} - \eta_{\nu\sigma}M_{\mu\lambda} +
\eta_{\nu\lambda}M_{\mu\sigma} - \eta_{\mu\lambda}M_{\nu\sigma}
\right)\;,\label{a1}\\
&&[\;M_i,P_j\;] = i\epsilon_{ijk}P_k\;,\label{a2}\\
&&[\;N_{i},P_j\;] = i\delta_{ij} {\rm
e}^{\frac{P_0}{2\kappa}}\left[\frac{\kappa}{2}\left(1-{\rm
e}^{-\frac{2P_0}{\kappa}}\right) + \frac{1}{2\kappa}{\rm
e}^{-\frac{P_0}{\kappa}} \vec{P}^2 \right] - \frac{i}{2\kappa}{\rm
e}^{-\frac{P_0}{2\kappa}}P_iP_j \;, \label{a3}\\
&&[\;M_{i},P_{0}\;] = 0 \;\;,\;\;\;\;[\;N_{i},P_{0}\;] = i{\rm
e}^{-\frac{P_0}{2\kappa}}P_i\;,\label{a4}\\
&&[\;P_{\mu},P_{\nu}\;] = 0\;,\label{a5}
\end{eqnarray}

b) coalgebraic sector
\begin{eqnarray}
&&\Delta(M_i) = M_i\otimes 1 + 1\otimes
M_i\;,\label{c1}\\
&&\Delta(N_i) = N_i\otimes 1 + {\rm e}^{-\frac{P_0}{\kappa}}\otimes
N_i + \frac{1}{\kappa}\epsilon_{ijk}{\rm
e}^{-\frac{P_0}{2\kappa}}P_j\otimes M_k\;,\label{c2}\\
&&\Delta(P_0) = P_0\otimes 1 + 1\otimes
P_0\;,\label{c3}\\
&&\Delta(P_i) = P_i\otimes {\rm e}^{\frac{P_0}{2\kappa}} + {\rm
e}^{-\frac{P_0}{2\kappa}}\otimes P_i\;,\label{c4}
\end{eqnarray}

c) coinverses (antipodes)
\begin{eqnarray}
&&S(M_i) = -M_i\;\;,\;\;S(N_i) = - {\rm e}^{\frac{P_0}{\kappa}}N_i +
\epsilon_{ijk}{\rm e}^{\frac{P_0}{2\kappa}}P_jM_k \;,\label{anti1}\\
&&~~~~~~~~~~~~~~~~~~~~~~~ S(P_i) = -P_i\;\;,\;\;S(P_0) = -P_0\;.
\label{anti2}
\end{eqnarray}
The $\kappa$-Poincare algebra has two $\kappa$-deformed Casimirs
describing mass and spin. The deformed bilinear mass Casimir looks
as follows
\begin{equation}
C_{\kappa}^2 (P_\mu) = C_{\kappa}^2 (\vec{P},P_0) = \left
(2\kappa\sinh \left(\frac{P_0}{2\kappa}\right)\right)^2 -
\vec{P}^2\;. \label{casimir}
\end{equation}
Because in any basis $C_{\kappa}^2 (P_\mu) = C_{\kappa}^2
(S(P_\mu))$ we see from (\ref{anti2}), (\ref{casimir}) that any
solution with $P_0>0$ has its "antiparticle counterpart" with $P_0
\to -P_0$.

The $\kappa$-deformed Minkowski space is defined as the translation
sector of dual $\kappa$-Poincare group \cite{lit13}, \cite{lit14}
with algebraic relations (\ref{minkowski}) and the classical
coproduct
\begin{equation}
 \Delta(\hat{x}_\mu) =
\hat{x}_\mu \otimes 1 + 1\otimes \hat{x}_\mu \;. \label{cominkowski}
\end{equation}

In order to introduce the noncommutative field operators one should
define the $\kappa$-deformed noncommutative plane waves. There are
possible various choices related with ordering ambiguity of the time
and space components of the plane waves (see e.g. \cite{lit22}), as
well as one can introduce the nonlinear transformations of the
fourmomentum variables (see e.g. \cite{lit23}). We shall choose the
formula
\begin{equation}
\vdots {\rm e}^{ip_\mu \hat{x}^\mu}\vdots ={\rm
e}^{\frac{i}{2}p_0{\hat x}^{0}}{\rm e}^{ip_i{\hat x}^{i}} {\rm
e}^{\frac{i}{2}p_0{\hat x}^{0}} \;, \label{exp}
\end{equation}
with simple Hermitean conjugation property
\begin{equation}
\left(\vdots {\rm e}^{ip_\mu \hat{x}^\mu}\vdots \right)^\dag =
\vdots {\rm e}^{-ip_\mu \hat{x}^\mu}\vdots \;. \label{property}
\end{equation}
From (\ref{minkowski}) and (\ref{exp}) follows the multiplication
rule
\begin{equation}
\vdots {\rm e}^{ip_\mu \hat{x}^\mu}\vdots \cdot \vdots {\rm
e}^{iq_\mu \hat{x}^\mu}\vdots = \vdots {\rm e}^{i\Delta_\mu(p,q)
\hat{x}^\mu}\vdots\;, \label{rule}
\end{equation}
where the fourvector $\Delta_\mu$ is determined by the coproduct
(\ref{c4})
\begin{equation}
\Delta_\mu(p,q) = \left(\Delta_0 = p_0 + q_0,~ \Delta_i = p_i{\rm
e}^{\frac{q_0}{2\kappa}} + q_i{\rm
e}^{-\frac{p_0}{2\kappa}}\right)\;. \label{delta}
\end{equation}
If we use the relation between two choices of noncommutative plane
waves
\begin{equation}
\vdots {\rm e}^{ip_\mu \hat{x}^\mu}\vdots = ~:{\rm
e}^{i\tilde{p}_\mu \hat{x}^\mu}:~ \equiv {\rm e}^{i\tilde{p}_i
\hat{x}^i}{\rm e}^{i\tilde{p}_0 \hat{x}^0}\;, \label{relation}
\end{equation}
where ${\tilde p}_\mu = ({\tilde p}_0 = p_0,~{\tilde p}_i={\rm
e}^{-\frac{p_{0}}{2\kappa}} p_i)$, the known formulae of the
$\kappa$-deformed bicovariant differential calculus \cite{lit24}
provide
\begin{equation}
\hat{\partial}_i~\vdots {\rm e}^{ip_\mu \hat{x}^\mu}\vdots = {\rm
e}^{\frac{p_0}{2\kappa}}p_i~\vdots {\rm e}^{ip_\mu
\hat{x}^\mu}\vdots\;, \label{calculus}
\end{equation}
and
\begin{equation}
\hat{\partial}_0~\vdots {\rm e}^{ip_\mu \hat{x}^\mu}\vdots = \left
[\kappa\left ({\rm e}^{\frac{p_0}{2\kappa}} - 1 \right ) +
\frac{1}{2\kappa}C^2_{\kappa}(p_\mu) \right ]~\vdots {\rm e}^{ip_\mu
\hat{x}^\mu}\vdots\;. \label{calculus1}
\end{equation}
Consequently, from (\ref{calculus}), (\ref{calculus1}) follows that
(see also \cite{lit16}, \cite{lit24}, \cite{lit22})
\begin{equation}
\hat{\Box}~\vdots {\rm e}^{ip_\mu \hat{x}^\mu}\vdots =
C^2_{\kappa}(p_\mu)\left (1 -
\frac{C^2_{\kappa}(p_\mu)}{4\kappa^2}\right ) \vdots {\rm e}^{ip_\mu
\hat{x}^\mu}\vdots\;. \label{kg}
\end{equation}

In the following section we shall apply the formulae
(\ref{exp})-(\ref{kg}) to the description of free quantum
noncommutative fields.

\section{Free $\kappa$-deformed quantum fields: new star product and microcausality.}

We shall describe the $\kappa$-deformed quantum scalar free field on
noncommutative Minkowski space (\ref{minkowski}) by the following
$\kappa$-deformed Fourier transform
\begin{equation}
{\hat \phi} ({\hat x}) = \frac{1}{(2\pi)^{\frac{3}{2}}} \int d^4p\;
A(p_0,\vec{p})\;\delta \left(C^2_{\kappa} (\vec{p},p_0) -
M^2\right)\vdots {\rm e}^{ip_\mu \hat{x}^\mu}\vdots\;. \label{field}
\end{equation}
As follows from the formula (\ref{kg}) the free fields satisfying
noncommutative Klein-Gordon equation can be described as a
superposition of two fields (\ref{field}) with two
$\kappa$-dependent masses - one physical, and second describing
ghost field (see e.g. \cite{lit16}).

We recall that in the quantum field (\ref{field}) the noncommutative
structures follows from two sources:

i) The presence of noncommutative Minkowski space coordinates
$\hat{x}_{\mu}$ (see (\ref{minkowski})),

ii) The noncommutativity of the Fourier modes $A(p_0,\vec{p})$.

If we solve the mass-shell condition $C^2_{\kappa} - M^2 = 0$ one
obtains
\begin{equation}
p_0^{\pm} = \pm\omega_\kappa(\vec{p})\;\;,\;\;\omega_\kappa(\vec{p})
= 2\kappa~{\rm arcsinh}\left (\frac{\sqrt{\vec{p}^2 +M^2}}{2\kappa}
\right)\;. \label{dis}
\end{equation}
Using the relation
\begin{equation}
\delta \left(C^2_{\kappa} (\vec{p},p_0) - M^2\right)  =
\frac{1}{2\Omega_{+}(\vec{p})}\delta(p_0 - p_0^{+}) +
\frac{1}{2\Omega_{-}(\vec{p})}\delta(p_0 - p_0^{-})\;, \label{rel}
\end{equation}
where\footnote{We choose the factor $\Omega_{\kappa} (\vec{p})$
having the standard $\kappa \to \infty$ limit equal to
$\omega(\vec{p})$.}
\begin{equation}
2{\Omega_{\pm}(\vec{p})} = \left|\frac{\partial}{\partial
p_0}C^2_{\kappa} (\vec{p},p_0) \right|_{p_0 = p_0^{\pm}} = 2\kappa
\sinh\left(\frac{\omega_{\kappa}(\vec{p})}{\kappa}\right) \equiv
2\Omega_{\kappa} (\vec{p})\;, \label{omega}
\end{equation}
and applying the decomposition into positive $(p_0 =
\omega_\kappa(\vec{p}))$ and negative $(p_0
=-\omega_\kappa(\vec{p}))$ frequency parts
\begin{equation}
{\hat \phi}({\hat x}) = {\hat \phi}_{+}({\hat x}) + {\hat
\phi}_{-}({\hat x})\;, \label{decom}
\end{equation}
we get
\begin{equation}
{\hat\phi}_{\pm} ({\hat x}) = \frac{1}{(2\pi)^{\frac{3}{2}}} \int
\frac{d^3\vec{p}}{2\Omega_{\kappa} (\vec{p})}\;
A(\pm\omega_{\kappa}(\vec{p}),\pm \vec{p})\;  \vdots {\rm e}^{\pm
ip_\mu \hat{x}^\mu}\vdots |_{p_0 = \omega_{\kappa}(\vec{p})}\;.
\label{fieldpm}
\end{equation}
In order to obtain the $\kappa$-deformed real scalar field (i.e. we
identify the particles and antiparticles)
\begin{equation}
\left( {\hat \phi}_{\pm}({\hat x})\right)^\dag =
{\hat\phi}_{\mp}({\hat x}) \;, \label{real}
\end{equation}
one should assume that
\begin{equation}
A^\dag(\pm\omega_{\kappa}(\vec{p}),\pm \vec{p})  =
A(\mp\omega_{\kappa}(\vec{p}),\mp \vec{p}) \;. \label{ass}
\end{equation}

In standard quantum field theory, if $\kappa \rightarrow \infty$,
the creation and annihilation operators $(\omega_{\infty}(\vec{p}) =
\omega(\vec{p}) = \sqrt{\vec{p}^{\ 2} + M^2})$
\begin{eqnarray}
a ({p}) = A(\omega(\vec{p}),\vec{p})\;\;,\;\; a^\dag ({p}) =
A(-\omega(\vec{p}),-\vec{p}) \;, \label{creation}
\end{eqnarray}
are quantized as follows
\begin{equation}
[\;a(p),a(q)\;] =  [\;a ^\dag (p),a^\dag (q)\;] =
0\;\;,\;\;[\;a^\dag (p),a (q)\;] =  2\omega(\vec{p})\delta^{(3)}
(\vec{p}-\vec{q})\;. \label{stanccr}
\end{equation}
The main question which now will be considered is how looks the
$\kappa$-deformation of the relations (\ref{stanccr}) describing
$\kappa$-deformed quantum free fields.

In order to represent the algebra of noncommutative fields on
$\kappa$-Minkowski space (\ref{minkowski}) by the fields on
classical space-time, we shall introduce the homomorphic
$\star$-product multiplication by means of the Weyl map $\vdots {\rm
e}^{ip_\mu\hat{x}^\mu}\vdots \rightarrow {\rm e}^{ip_\mu{x}^\mu}$
reproducing the relation (\ref{rule}), i.e.
\begin{equation}
\vdots {\rm e}^{ip_\mu\hat{x}^\mu}\vdots \cdot \vdots{\rm
e}^{iq_\mu\hat{x}^\mu}\vdots  \leftrightarrow {\rm
e}^{ip_\mu{x}^\mu}\star {\rm e}^{iq_\mu{x}^\mu}= {\rm
e}^{i(p_0+q_0)x^0 + i\Delta_i(\vec{p},\vec{q})x^i}\;. \label{point1}
\end{equation}
Now we shall introduce the new step - we assume that the
star-product of the quantum free fields (\ref{field}) affects also
the mass-shell conditions. We propose the following new star-product
$\star_{\kappa}$ of two free $\kappa$-deformed fields
\begin{eqnarray}
&&{\hat \phi}(\hat{x})\cdot {\hat\phi}(\hat{x})\leftrightarrow
\varphi(x)\star_{\kappa} \varphi(x) = \frac{1}{(2\pi)^{3}} \int
d^4p\int d^4q\; A(p_0,\vec{p})A(q_0,\vec{q})~{\rm e}^{i(p_0+q_0)x^0
+ i\Delta_i(\vec{p},\vec{q})x^i}\nonumber\\
&&~~~~~~~~~~~~~~~~~~~~~~~~~~~~~~~~~~~~~~~~ \cdot\delta (
C^2_{\kappa} (\vec{p}{\rm e}^{\frac{q_0}{2\kappa}},p_0) - M^2)~
\delta ( C^2_{\kappa} (\vec{q}{\rm e}^{-\frac{p_0}{2\kappa}},q_0) -
M^2 )\;. \label{newstar}
\end{eqnarray}
In order to consider the bilocal product of noncommutative fields we
extend the Weyl map (\ref{point1}) as follows (see also
\cite{lit18})
\begin{equation}
\vdots {\rm e}^{ip_\mu\hat{x}^\mu}\vdots \cdot\vdots {\rm
e}^{iq_\mu\hat{y}^\mu}\vdots \leftrightarrow{\rm e}^{ip_\mu{x}^\mu}
\star {\rm e}^{iq_\mu{y}^\mu}
= {\rm e}^{i(p_0x^0+q_0y^0) +
(p_i{\rm e}^{\frac{q_0}{2\kappa}}x^i + q_i{\rm
e}^{-\frac{p_0}{2\kappa}}y^i)}\;. \label{twopoint}
\end{equation}
From (\ref{twopoint}) we obtain by differentiation the braid
relations between two copies of $\kappa$-Minkowski space
\begin{eqnarray}
&&x_0 \star y_0 = x_0 y_0\;\;,\;\;~~~~~~~~~~~~y_0 \star x_0 = y_0
x_0\;\;~~~~~~~~~~~~~~~\Rightarrow ~~~~\;\;
[\;{x}_{0},{y}_{0}\;]_{\star} =
0\;,\;\;\;\;\;\;\;\;\;\;\;\;\; \label{twocom1}\\
&&x_i \star y_j = x_i y_j\;\;,\;\;~~~~~~~~~~~~~y_j \star x_i = y_j
x_i\;\;~~~~~~~~~~~~~~~~\Rightarrow ~~~~\;\;
[\;{x}_{i},{y}_{j}\;]_{\star} =
0\;,\;\;\;\;\;\;\;\;\;\;\;\;\;\;\; \label{twocom4}\\
&&x_0 \star y_i = \frac{i}{2\kappa}y_i +y_ix_0\;\;,\;\;~~~y_i \star
x_0 = -\frac{i}{2\kappa}y_i +y_ix_0\;\;~~~\Rightarrow ~~~~\;\;
[\;{x}_{0},{y}_{i}\;]_{\star} = \frac{i}{\kappa}y_i\;,
\label{twocom2}\\
&&x_i \star y_0 = -\frac{i}{2\kappa}x_i +x_iy_0\;\;,\;\;y_0 \star
x_i = \frac{i}{2\kappa}x_i +x_iy_0\;\;~~~~~~\Rightarrow~~~~ \;\;
[\;{y}_{0},{x}_{i}\;]_{\star} = \frac{i}{\kappa}x_i\;.
\label{twocom3}
\end{eqnarray}
For the bilocal product of free $\kappa$-deformed quantum fields we
extend (\ref{newstar}) in the following way
\begin{eqnarray}
&&{\hat \phi}(\hat{x})\cdot {\hat \phi}(\hat{y})\leftrightarrow
\varphi(x)\star_{\kappa} \varphi(y) = \frac{1}{(2\pi)^{3}} \int
d^4p\;\int d^4q\; A(p_0,\vec{p})A(q_0,\vec{q})\ {\rm
e}^{i(p_0x^0+q_0y^0) + (p_i{\rm e}^{\frac{q_0}{2\kappa}}x^i +
q_i{\rm
e}^{-\frac{p_0}{2\kappa}}y^i)}\nonumber \\
&&~~~~~~~~~~~~~~~~~~~~~~~~~~~~~~~~~~~~~~~~~~~~~ \cdot\;\delta (
C^2_{\kappa} (\vec{p}{\rm e}^{\frac{q_0}{2\kappa}},p_0) - M^2)~
\delta ( C^2_{\kappa} (\vec{q}{\rm e}^{-\frac{p_0}{2\kappa}},q_0) -
M^2 )\;.\label{newstar1}
\end{eqnarray}
We stress that the star product $\star_{\kappa}$ is different from
the standard one, generated by the formula (\ref{twopoint}). It
introduces new multiplication rule of two free fields, which is
characterized by additional shifts $\vec{p} \to \vec{p}{\rm
e}^{\frac{q_0}{2\kappa}}$ and $\vec{q} \to \vec{q}{\rm
e}^{-\frac{p_0}{2\kappa}}$ of the threemomentum variables occuring
in the arguments of the mass-shell deltas. The prescription
(\ref{newstar}) is defined only for the free fields, with
fourmomenta restricted by $\kappa$-deformed mass-shell condition,
and in interacting case can be used only for the description of
$\kappa$-deformed perturbative expansions.

We shall relate the $\star_{\kappa}$-multiplication (\ref{newstar1})
with the $\kappa$-deformed statistics, described by the
$\kappa$-deformation of the relations (\ref{stanccr}). Let us
introduce new variables
\begin{equation}
{\cal{P}}_{0} = p_{0}\;\;\;,\;\;\;{\cal{P}}_{i} = p_{i}{\rm
e}^{\frac{q_0}{2\kappa}}\;\;\;,\;\;\;{\cal{Q}}_{0} =
q_{0}\;\;\;,\;\;\;{\cal{Q}}_{i} = q_{i}{\rm
e}^{-\frac{p_0}{2\kappa}}\;, \label{variables}
\end{equation}
in the momentum integrals in (\ref{newstar1}). One gets
\begin{eqnarray}
&&\varphi({x})\star_{\kappa} \varphi({y}) = \frac{1}{(2\pi)^{3}}
\int d^4{\cal{P}}\;\int d^4{\cal{Q}}\; {\rm e}^{\frac{3({\cal{P}}_0
- {\cal{Q}}_{0})}{2\kappa}} A({\cal{P}}_0,{\rm e}^{-\frac{{\cal
Q}_0}{2\kappa}}{{\cal{P}}}_i) A({\cal{Q}}_0,{\rm e}^{\frac{{\cal
P}_0}{2\kappa}}{{\cal{Q}}}_i)~
{\rm e}^{i({\cal{P}}_\mu x^\mu+{\cal{Q}}_\mu y^\mu)} \;\nonumber\\
&&~~~~~~~~~~~~~~~~~~~~~~~~~~~~~~~~~~~~~~~~~~~~~~~ \cdot\delta (
C^2_{\kappa} (\vec{{\cal{P}}},{\cal{P}}_0) - M^2)~ \delta (
C^2_{\kappa} (\vec{{\cal{Q}}},{\cal{Q}}_0) - M^2 )\;.
\label{newstar2}
\end{eqnarray}
Using the formula (\ref{newstar2}) after the exchange $x
\leftrightarrow y$ we have
\begin{eqnarray}
&&[\;\varphi (x),\varphi (y)\;]_{\star_{\kappa}} =
\frac{1}{(2\pi)^{3}} \int d^4{\cal{P}}\;\int d^4{\cal{Q}}\; \left
[{\rm e}^{\frac{3({\cal{P}}_0 - {\cal{Q}}_{0})}{2\kappa}}
A\left({\cal{P}}_0,{\rm e}^{-\frac{{\cal{Q}}_0}{2\kappa}}
\vec{{\cal{P}}} \right) A\left({\cal{Q}}_0,{\rm
e}^{\frac{P_0}{2\kappa}}\vec{{\cal{Q}}} \right) - \right .\nonumber\\
&&~~~~~~~~~~~~~~~~~~~~~~~~~~~~~~~~~~~~~~~~~~~~~~~\left . - {\rm
e}^{\frac{-3({\cal{P}}_0 - {\cal{Q}}_{0})}{2\kappa}}
A\left({\cal{Q}}_0,{\rm e}^{-\frac{{\cal{P}}_0}{2\kappa}}
\vec{{\cal{Q}}} \right) A\left({\cal{P}}_0,{\rm
e}^{\frac{{\cal{Q}}_0}{2\kappa}}\vec{{\cal{Q}}} \right)\right ]
\nonumber\\
&&~~~~~~~~~~~~~~~~~~~~~~~~~~~~~~~~~~~~\cdot{\rm e}^{i({\cal{P}}_\mu
x^\mu+{\cal{Q}}_\mu y^\mu)}\delta ( C^2_{\kappa}
(\vec{{\cal{P}}},{\cal{P}}_0) - M^2)~ \delta ( C^2_{\kappa}
(\vec{{\cal{Q}}},{\cal{Q}}_0) - M^2 )\;. \label{commu}
\end{eqnarray}

Now, we introduce the $\kappa$-deformed creation and annihilation
operators as follows
\begin{eqnarray}
a_\kappa ({p}) = A(\omega_{\kappa}(\vec{p}),\vec{p})\;\;\;,\;\;\;
a_{\kappa}^\dag ({p}) = A(-\omega_{\kappa}(\vec{p}),-\vec{p}) \;.
\label{creani}
\end{eqnarray}
We also define the following new $\kappa$-deformed multiplication
for the operators (\ref{creani})\footnote{The formulae
(\ref{multi0})-(\ref{multi3}) differ from the ones studied in detail
in \cite{lit18} by the functional normalization factor
$F_\kappa^{(2)}(\vec{p},\vec{q})$; see however that (\ref{multi0})
is a special case of the general formula (29) from \cite{lit18}.}
\begin{eqnarray}
a_\kappa ({p})\circ
a_\kappa({q})&=&F_\kappa^{(2)}(p_0,q_0)\;a_\kappa \left(p_0,{\rm
e}^{-\frac{q_0}{2\kappa}}\vec{p} \right) a_\kappa \left(q_0,{\rm
e}^{\frac{p_0}{2\kappa}}\vec{q} \right)\;,\label{multi0}\\
a_\kappa^\dag ({p})\circ a_\kappa^\dag ({q})&=&
F_\kappa^{(2)}(-p_0,-q_0)\;a_\kappa^\dag \left(p_0,{\rm
e}^{\frac{q_0}{2\kappa}}\vec{p} \right) a_\kappa^\dag \left(q_0,{\rm
e}^{-\frac{p_0}{2\kappa}}\vec{q} \right)\;,\label{multi1}
\end{eqnarray}
and
\begin{eqnarray}
a_\kappa^\dag ({p})\circ a_\kappa({q})&=&
F_\kappa^{(2)}(-p_0,q_0)\;a_\kappa^\dag \left(p_0,{\rm
e}^{-\frac{q_0}{2\kappa}}\vec{p} \right) a_\kappa \left(q_0,{\rm
e}^{-\frac{p_0}{2\kappa}}\vec{q}
\right)\;,\label{multi2}\\
a_\kappa ({p})\circ a_\kappa^\dag({q})&=& F_\kappa^{(2)}(p_0,-q_0)\;
a_\kappa\left(p_0,{\rm e}^{\frac{q_0}{2\kappa}}\vec{p} \right)
a_\kappa^\dag \left(q_0,{\rm e}^{\frac{p_0}{2\kappa}}\vec{q}
\right)\;,\label{multi3}
\end{eqnarray}
where $p_0 = \omega_{\kappa}(\vec{p})$, $q_0 =
\omega_{\kappa}(\vec{q})$ and $F_\kappa^{(2)}(p_0,q_0) = {\rm
e}^{{{\frac{3}{2\kappa}(p_0 - q_0)}}}$\footnote{The derivation of
the relations (\ref{multi0})-(\ref{multi3}) is based on two steps.
Firstly, we consider the consistency with non-Abelian addition law
(\ref{c4}) for the threemomenta. Using the Hopf-algebraic formula
$$
\hspace{3.5cm}P_i \triangleright \left(a(p)a(q)\right) = \left(
\Delta_{(1)} \triangleright a(p)\right)\cdot \left(\Delta_{(2)}
\triangleright a(q)\right)\;,\hspace{4.0cm} (*)$$ where we assume
that (see also \cite{lit12}, \cite{ArMar}) $$ P_i \triangleright
a(p) = p_ia(p)\;,$$ we get $$P_i \triangleright
\left(a(p)a(q)\right) = p_i^{(1+2)} a(p)a(q)\;,$$ where
$$p_i^{(1+2)} =
{\rm{e}}^{\frac{q_0}{2\kappa}}p_i +
{\rm{e}}^{-\frac{p_0}{2\kappa}}q_i\;. $$ We should modify
threemomenta of the exchanged oscillators $(\vec{p} \rightarrow
\vec{p}_{\kappa},\vec{q} \rightarrow \vec{q}_{\kappa})$ in such a
way that $$\hspace{4.5cm}P_i\triangleright
\left(a(q_{\kappa})a(p_{\kappa})\right) = p_i^{(1+2)}
a(q_{\kappa})a(p_{\kappa})\;.\hspace{4.5cm} (**)$$ One gets, using
$(*)$, that
$$\vec{p}_{\kappa} =
{\rm{e}}^{\frac{q_0}{\kappa}}\vec{p}\;\;,\;\;\vec{q}_{\kappa} =
{\rm{e}}^{-\frac{p_0}{\kappa}}\vec{q}\;.
$$In order to obtain classical threemomenta addition law we perform the
second step: we change in the relations $(*)$ and $(**)$ the
variables $(\vec{p},\vec{q}) \rightarrow
({\rm{e}}^{-\frac{q_0}{2\kappa}}\vec{p},{\rm{e}}^{\frac{p_0}{2\kappa}}\vec{q})$,
what leads to the classical formula for the two-particle
threemomenta. In such a way we obtain relation (\ref{multi0}); the
derivation of the relations (\ref{multi1})-(\ref{multi3}) is
analogues provided that $$P_{\mu}\triangleright a^{\dag}(p) =
-p_{\mu}a^{\dag}(p)\;.$$}. Our basic postulate is the following set
of $\kappa$-deformed creation and annihilation operators
$([\;A,B\;]_\circ := A\circ B - B\circ A)$
\begin{equation}
[\;a_\kappa(p),a_\kappa(q)\;]_{\circ} =  [\;a_\kappa ^\dag
(p),a_\kappa^\dag (q)\;]_{\circ} = 0 \;\;\;,\;\;\; [\;a_\kappa^\dag
(p),a_\kappa (q)\;]_{\circ} =  2\Omega_\kappa(\vec{p})\delta^{(3)}
(\vec{p}-\vec{q})\;, \label{kappaccr2}
\end{equation}
where the $\kappa$-deformation of standard formulae (\ref{stanccr})
is contained in the deformed multiplication rules
(\ref{multi0})-(\ref{multi3}).

Let us introduce the $\circ$-product of two $\kappa$-deformed scalar
fields as follows
\begin{eqnarray}
&&\varphi(x)\circ \varphi(y) = \frac{1}{(2\pi)^{3}} \int d^4{\cal
P}\int d^4{\cal Q}\; A({\cal P}_0,\vec{{\cal P}})\circ A({\cal
Q}_0,\vec{{\cal Q}})\cdot~{\rm e}^{i({\cal P}x+{\cal
Q}y)}~~~~~~~~~~~~~~~~~~~~~~~~~~~~~~ \nonumber
\\
&&~~~~~~~~~~~~~~~~~~~~~~~~~~~~~~~~~~~~~~~~~~~~~~~~~~~~~\cdot \delta
( C^2_{\kappa} (\vec{{\cal P}},{\cal P}_0) - M^2)~ \delta (
C^2_{\kappa} (\vec{{\cal Q}},{\cal Q}_0) - M^2 )\;, \label{circpro}
\end{eqnarray}
where the product $A(p_0,\vec{p})\circ A(q_0,\vec{q})$ is determined
via formula (\ref{creani}) by four definitions
(\ref{multi0})-(\ref{multi3}). If we introduce the formula inverse
to the one given by (\ref{variables}), the nonlinear change of the
momentum variables $({\cal P},{\cal Q}) \to (p,q)$ leads to the
identity
\begin{equation}
\varphi(x)\star_{\kappa} \varphi(y) = \varphi(x)\circ \varphi(y)\;.
\label{identity}
\end{equation}

We would like to recall that the relation (\ref{identity}) for
canonical $\theta_{\mu\nu}$-deformation ($\theta_{\mu\nu} = $ const)
is known (see \cite{lit12}, eq. (3.8)). One can say that if we
postulate the $\kappa$-statistics using the relation
(\ref{kappaccr2}), we can derive the $\star_{\kappa}$-multiplication
given by the formula (\ref{newstar}) from the relation
(\ref{identity}).

We see therefore that our new star product $\star_{\kappa}$ can be
equivalently described by the $\circ$-multiplication describing
$\kappa$-deformation of the oscillator algebra. After simple
calculation one can show that\footnote{See also the relations
(\ref{micro1}), (\ref{pauli}).}
\begin{eqnarray}
&&[\;\varphi (x),\varphi (y)\;]_{\star_{\kappa}} =[\;\varphi
(x),\varphi (y)\;]_{\circ} = \frac{-i}{(2\pi)^{3}} \int
\frac{d^3p}{\Omega_\kappa(\vec{p})}\sin(\omega_\kappa(\vec{p})(x_0-y_0)){\rm
e}^{i\vec{p}(\vec{x}-\vec{y})} =
\nonumber\\
&&~~~~~~~~~~~~~~~~~~~~~~~~~~~~~~~~~~~~~~~~~~~~~~~~~~~~~
~~~~~=\frac{1}{i}\Delta_\kappa (x-y;M^2)\;. \label{microcausality1}
\end{eqnarray}

It should be noted that the commutator (\ref{microcausality1}) was
firstly obtained in \cite{lit19} by using "naive" deformation of the
free quantum K-G fields (the "ad hoc" insertion of $\kappa$-deformed
mass-shell condition).

Let us consider the equal time properties of the $\kappa$-deformed
Pauli-Jordan commutator function. It follows from
(\ref{microcausality1}) that
\begin{equation}
\Delta_\kappa (x-y;M^2)|_{x_0 = y_0} = 0
\;\;\;\Leftrightarrow\;\;\;[\;\varphi (x_0,\vec{x}),\varphi
(x_0,\vec{y})\;]_{\star_{\kappa}} = 0\;, \label{property1}
\end{equation}
\begin{equation}
{\partial_0}\Delta_\kappa (x-y;M^2)|_{x_0 = y_0}= [\;{\partial
_0}\varphi (x_0,\vec{x}),\varphi (x_0,\vec{y})\;]_{\star_{\kappa}} =
\delta_{\kappa}^{(3)}(\vec{x}-\vec{y})\;, \label{property2}
\end{equation}
where
\begin{equation}
\delta_{\kappa}^{(3)}(\vec{x}-\vec{y}) = \frac{1}{(2\pi)^{3}} \int
{d^3\vec{p}}\frac{\omega_{\kappa}(\vec{p}^2)}{\Omega_\kappa(\vec{p}^2)}{\rm
e}^{i\vec{p}(\vec{x}-\vec{y})}
\;\;\;{~}_{\overrightarrow{{\;\;\;\kappa \rightarrow
\infty}\;\;\;}}\;\;\;\delta^{(3)}(\vec{x}-\vec{y})\;,
\label{kappadelta}
\end{equation}
is the $\kappa$-deformed nonlocal counterpart of standard Dirac
delta function. It appears however that if we define "quantum" time
derivative
\begin{equation}
\partial_0^{\kappa} \equiv i\kappa \sinh\left (\frac{1}{i}\frac{\partial_0}{\kappa}\right)
= \kappa \sin \left
(\frac{\partial_0}{\kappa}\right)\;\;\;{~}_{\overrightarrow{{\;\;\;\kappa
\rightarrow \infty}\;\;\;}}\;\;\;
\partial_0\;, \label{qder}
\end{equation}
one gets the local expression for any value of $\kappa$
\begin{equation}
\partial_0^{\kappa}\Delta_\kappa (x-y;M^2)|_{x_0 = y_0}=
\delta^{(3)}(\vec{x}-\vec{y})\;. \label{delta}
\end{equation}
The formulae (\ref{property2})-(\ref{delta}) describe the equal time
relations representing the $\kappa$-causality of $\kappa$-deformed
quantum fields. It should be stressed that such results require the
$\kappa$-deformed algebra (\ref{multi0})-(\ref{multi3}) of the field
oscillators.

\section{The algebra of $\kappa$-deformed bosonic creation and annihilation operators.}

In order to study the full algebra of creation and anihilation
operators one should extend the relations
(\ref{multi0})-(\ref{multi3}) to the $\circ$-product of arbitrary
polynomials of creation and annihilation operators. Because due to
the relation (\ref{creani}) one can describe the relations
(\ref{multi1})-(\ref{multi3}) by suitable extension of the relation
(\ref{multi0}) to negative energies, further we shall consider only
the products of the creation operators $a_\kappa(p)$ with arbitrary
values of $p$ (sign of $p_0$ is not specified).

The formula (\ref{multi0}) can be extended to an arbitrary product
of $n$ $\kappa$-deformed creation oscillators in the following way
\begin{eqnarray}
&&a_\kappa({p}^{(1)} )\circ \cdots \circ a_\kappa({p}^{(n)} ) =
F_{\kappa}^{(n)}(p_0^{(1)},\ldots,p_0^{(n)})a_\kappa\left(p_0^{(1)},
\chi_{n}^{(1)}(p_0^{(1)},\ldots,p_0^{(n)})\vec{p}^{~(1)}\right
)\cdots
 \nonumber\\
&&~~~~~~~~~~~~~~\cdots
a_\kappa\left(p_0^{(k)},\chi_{n}^{(k)}(p_0^{(1)},\ldots,
p_0^{(n)})\vec{p}^{~(k)} \right)\cdots
a_\kappa\left(p_0^{(n)},\chi_{n}^{(n)}(p_0^{(1)},\ldots,
p_0^{(n)})\vec{p}^{~(n)} \right)\;, \label{n}
\end{eqnarray}
where
\begin{equation}
\chi_{n}^{(k)}(p_0^{(1)},\ldots, p_0^{(n)}) = \exp \frac{1}{2\kappa}
\left( {\sum\limits_{j=1}^{k-1}}p_0^{(j)} -
{\sum\limits_{j=k+1}^{n}}p_0^{(j)}\right)\;, \label{cchi}
\end{equation}
\begin{equation*}
F_{\kappa}^{(n)}(p_0^{(1)},\ldots,p_0^{(n)}) =  \left[ ~
{\prod\limits_{k=1}^{n}} \chi_{n}^{(k)}(p_0^{(1)},\ldots, p_0^{(n)})
~\right]^3 = {\rm
exp}\left({\frac{3}{2\kappa}{\sum\limits_{k=1}^{n}(n+1-2k)}p_0^{(k)}}\right)
\;,\nonumber
\end{equation*}
or using more compact notation
\begin{equation}
a_\kappa({p}^{(1)} )\circ \cdots \circ a_\kappa({p}^{(n)} ) =
F_\kappa^{(n)}(p_0^{(1)},\ldots,p_0^{(n)})
a_\kappa\left(p_0^{(1)},\vec{{\cal P}}_{n}^{(1)} \right) \cdots
a_\kappa\left(p_0^{(n)},\vec{{\cal P}}_{n}^{(n)} \right)\;,
\label{n1}
\end{equation}
with
\begin{equation}
{\vec {\cal P}}_n^{(k)} =  \exp \frac{1}{2\kappa} \left(
{\sum\limits_{j=1}^{k-1}}p_0^{(j)} -
{\sum\limits_{j=k+1}^{n}}p_0^{(j)}\right) \vec{{p}}^{~(k)} =
\chi_{n}^{(k)}(p_0^{(1)},\ldots
p_0^{(n)})\vec{{p}}^{~(k)}\;.\label{calmom}
\end{equation}

Let us observe that the factors $\chi_{n}^{(k)}$ are determined by
the n-th iteration of the coproduct of the fourmomenta in the basis
with the basic coproducts (\ref{c3}), (\ref{c4})
\begin{equation}
\Delta^{(n)} (P_0) = {\sum\limits_{k=1}^{n}} ~{\underbrace{1\otimes
\cdots \otimes 1}_{k-1}}~ \otimes P_0 \otimes ~{\underbrace{1\otimes
\cdots\otimes 1}_{n-k}}\;, \label{nthco}
\end{equation}
\begin{equation}
\Delta^{(n)} (P_i) = {\sum\limits_{k=1}^{n}} ~{\underbrace{{\rm
e}^{\frac{P_0}{2\kappa}}\otimes \cdots \otimes {\rm
e}^{\frac{P_0}{2\kappa}}}_{k-1}}~ \otimes P_i \otimes
~{\underbrace{{\rm e}^{-\frac{P_0}{2\kappa}}\otimes \cdots\otimes
{\rm e}^{-\frac{P_0}{2\kappa}}}_{n-k}}\;. \label{nthco1}
\end{equation}

The relation (\ref{nthco1}) encodes the deformed addition law for
the threemomenta, but if we assume that (see also \cite{lit12},
\cite{ArMar})
\begin{equation}
P_{\mu}\rhd a_{\kappa}(p) = p_\mu a_{\kappa}(p)\;, \label{posc}
\end{equation}
we have (we use Sweedler notation for n-fold coproduct $\Delta^{(n)}
= \Delta_{(1)}^{(n)}\otimes\cdots\otimes \Delta_{(n)}^{(n)})$
\begin{eqnarray}
&&P_0 \triangleright(a_\kappa(p^{(1)})\circ \ldots \circ
a_\kappa(p^{(n)})) = F_{\kappa}^{(n)}\omega \left (\Delta^{(n)}(P_0)
\triangleright (a_\kappa({\cal P}_n^{(1)})\otimes \cdots \otimes
a_\kappa({\cal P}_n^{(n)})
)\right)=   \nonumber\\
&&~~~~~~~~~~~~=F_{\kappa}^{(n)}\omega \left (
\Delta_{(1)}^{(n)}(P_0) \triangleright a_\kappa({{\cal
P}}_n^{(1)})\otimes \cdots \otimes \Delta_{(n)}^{(n)}(P_0)
\triangleright a_\kappa({{\cal P}}_n^{(n)})
\right) = \nonumber\\
&&~~~~~~~~~~~~=
{\sum\limits_{j=1}^{n}}~p_0^{(j)}~a_\kappa(p^{(1)})\circ \ldots
\circ a_\kappa(p^{(n)})\;, \label{f1}
\end{eqnarray}
where $\omega (A\otimes B) = AB$, and
\begin{eqnarray}
&&\vec{P}\triangleright (a_\kappa(p^{(1)})\circ \ldots \circ
a_\kappa(p^{(n)})) = F_{\kappa}^{(n)}\omega\left (
\Delta^{(n)}(\vec{P}) \triangleright (a_\kappa({{\cal
P}}_n^{(1)})\otimes \cdots \otimes a_\kappa({{\cal P}}_n^{(n)})
)\right) =  \nonumber\\
&&~~~~~~~~~~~~~~~~~~=F_{\kappa}^{(n)}\omega \left (
\Delta_{(1)}^{(n)}(\vec{P}) \triangleright a_\kappa({{\cal
P}}_n^{(1)})\otimes \cdots \otimes \Delta_{(n)}^{(n)}(\vec{P})
\triangleright a_\kappa({{\cal P}}_n^{(n)})
\right) =\nonumber\\
&&~~~~~~~~~~~~~~~~~~= {\sum\limits_{j=1}^{n}}~\vec{{\cal
P}}_n^{(j)}\left (\chi_{n}^{(j)}(p_0^{(1)},\ldots, p_0^{(n)}) \right
)^{-1}~a_\kappa(p^{(1)})\circ \ldots \circ a_\kappa(p^{(n)}) =
\nonumber\\
&&~~~~~~~~~~~~~~~~~~=
{\sum\limits_{j=1}^{n}}~\vec{{p}}^{~(j)}~a_\kappa(p^{(1)})\circ
\ldots \circ a_\kappa(p^{(n)})\;. \label{f2}
\end{eqnarray}
We see therefore that the modification of the momentum arguments in
the $\kappa$-deformed oscillators (\ref{multi0})-(\ref{multi3})
exactly cancels non-Abelian factors in the quantum addition law,
governed by the coproduct (\ref{nthco1}).

Using the relations
(\ref{nthco}), (\ref{nthco1}) one can also explain the meaning of
our new star-product binary multiplication (\ref{newstar1}).
Let us introduce new set of threemomentum variables (compare with
(\ref{calmom}))
\begin{equation}\label{hatp}
{\hat{\vec{\cal P}}}^{(i)}_n = \chi^i_n (p_0^{(1)}, \ldots, p^{(n)}_0)^{-1}
\vec{p}^{~(i)} \; ,
\end{equation}
which describe the $i$-th particle contribution to the total
$n$-particle threemomentum, obtained from the coproduct
(\ref{nthco1}).
For $n=2$ one can  interpret the
threevectors ${\hat{\vec{\cal{P}}}}_2^{(1)}$,
 ${\hat{\vec{\cal{P}}}}_2^{(2)}$ as
describing the first and second particle threemomenta in a
two-particle state $a_{\kappa}(p_1)a_{\kappa}(p_2)|0>$, because
\begin{equation}
\vec{P} \triangleright (a_\kappa(p^{(1)})a_\kappa(p^{(2)})) = \omega
\left(\Delta^{(2)} \triangleright (a_\kappa(p^{(1)})\otimes
a_\kappa(p^{(2)}))\right) = \left({\hat{\vec{\cal{P}}}}_2^{(1)} +
{\hat{\vec{\cal{P}}}}_2^{(2)}\right)(a_\kappa(p^{(1)})a_\kappa(p^{(2)})) \;.
\label{atwopar}
\end{equation}
The deformation of the product of two mass-shell conditions in the
formula (\ref{newstar1}) can be described by the following
replacement  $(i=1,2)$
\begin{equation}
C^2_{\kappa}(p_0^{(i)},\vec{p}^{~(i)}) \to
C^2_{\kappa}(p_0^{(i)},{\hat{\vec{\cal{P}}}}_2^{(i)})\;, \label{replacment}
\end{equation}
i.e. we put the threemomenta
${\hat{\vec{\cal{P}}}}_2^{(1)}$, ${\hat{\vec{\cal{P}}}}^{(2)}_2$
  on
$\kappa$-deformed mass-shell.

The relation (\ref{replacment}) can be extended to the product of n
$\kappa$-deformed free fields containing product of n $\kappa$-deformed
mass-shell conditions. In a such case the relation (\ref{identity})
we extend as follows
\begin{equation}
\varphi(x_1) \star_{\kappa} \ldots \star_{\kappa} \varphi(x_n) =
\varphi(x_1) \circ \ldots \circ \varphi(x_n)\;, \label{ext}
\end{equation}
where the rhs of (\ref{ext}) is defined with the use of the relation
(\ref{n}). In order to define consistently the lhs of (\ref{ext}) we
should replace in the i-th mass-shell $\vec{p}^{~(i)} \to
{\hat{\vec{\cal{P}}}}_n^{(i)}$
 (see (\ref{hatp})).

The $\kappa$-deformed algebra of oscillators can be formulated in a
way consistent with the associativity property of the product
(\ref{n}). Indeed, one can define the product $\left(
a_\kappa\left({p}^{(1)} \right)\cdot\; \cdots \;\cdot
a_\kappa\left({p}^{(n)} \right) \right) \circ \left(
a_\kappa\left({q}^{(1)} \right)\cdot\; \cdots \;\cdot
a_\kappa\left({q}^{(m)} \right) \right)$ in such a way that for any
$n$, $m$ the relation
\begin{eqnarray}
&&\left( a_\kappa({p}^{(1)} )\circ \cdots \circ a_\kappa({p}^{(n)} )
\right) \circ \left( a_\kappa({q}^{(1)} )\circ \cdots \circ
a_\kappa({q}^{(m)} ) \right) =\nonumber\\
&&~~~~~~~~~~~~~~~~~~~~~~~~~~~~~~~~~~~~
 = \left( a_\kappa({p}^{(1)}
)\circ \cdots \circ a_\kappa({p}^{(n)}) \circ a_\kappa({q}^{(1)}
)\circ \cdots \circ a_\kappa({q}^{(m)}) \right)\;,
\label{associativity}
\end{eqnarray}
is valid. If we insert the relation (\ref{associativity}) into the
relation (\ref{ext}) we see that the star multiplication
$\star_{\kappa}$ of $\kappa$-deformed free fields is also
associative.

\section{$\kappa$-deformed Fock space, fourmomentum conservation la-ws and statistics.}

Let us introduce the normalized vacuum state in standard way
\begin{equation}
a_\kappa^\dag (p_0,\vec{p})|0> = 0\;\;\;,\;\;\; <0|0> = 1\;,
\label{vacuum}
\end{equation}
where $p_0 = \omega_{\kappa}(\vec{p})$. The one-particle state is
defined as follows\footnote{If we consider single oscillators and
one-particle states one can put $a_\kappa (p_0,\vec{p})\equiv a
(p_0,\vec{p})$, because the $\kappa$-deformation appears only if we
multiply the oscillators.}
\begin{equation}
|\vec{p}> =  a_\kappa(p_0,\vec{p})|0>\;, \label{one}
\end{equation}
and from (\ref{posc}) we get
\begin{equation}
P_{\mu}|\vec{p}> =  p_{\mu}|\vec{p}>\;. \label{pone}
\end{equation}
We define two-particle states in the following way
\begin{equation}
|\vec{p},\vec{q}> =  a_\kappa(p_0,\vec{p})\circ
a_\kappa(q_0,\vec{q})|0>\;. \label{two}
\end{equation}
By using (\ref{nthco}) and (\ref{nthco1}) we obtain\footnote{We
should stress that two quanta in (\ref{two}) are forming an
intertwined system and should not be consider as a superposition of
two independent modes. The intertwining effect puts each component
of 2-particle state off-shell. This footnote contains the answer to
the criticism presented in \cite{ArMar}.}
\begin{equation}
P_{\mu}|\vec{p},\vec{q}> =  (p_{\mu}+q_{\mu})|\vec{p},\vec{q}>\;.
\label{ptwo}
\end{equation}
From (\ref{stanccr}) follows the standard bosonic symmetry
\begin{equation}
|\vec{p},\vec{q}> = |\vec{q},\vec{p}>\;. \label{symmetryb}
\end{equation}
Taking into account associativity property of
$\kappa$-multiplication (see (\ref{associativity})) one can define
the n-particle state by the $\kappa$-deformed product of n
oscillators. We obtain
\begin{equation}
|\vec{p}^{~(1)},\ldots,\vec{p}^{~(k)},\ldots,\vec{p}^{~(n)}> =
a_\kappa(p^{(1)})\circ \ldots \circ a_\kappa(p^{(k)})\circ \ldots
\circ a_\kappa(p^{(n)})|0> \label{state}\,.
\end{equation}
Using the coassociative coproduct (\ref{nthco}) and (\ref{nthco1})
of n-th order, we can calculate from (\ref{f1}), (\ref{f2}) the
total momentum of the state (\ref{state}), which appears to be given
by classical formula
\begin{eqnarray}
&&P_\mu |\vec{p}^{~(1)},\ldots,\vec{p}^{~(k)},\ldots,\vec{p}^{~(n)}>
= \left [ \Delta^{(n)}(P_\mu) \triangleright (a_\kappa(p^{(1)})\circ
\ldots \circ a_\kappa(p^{(n)})
)\right]|0> = \nonumber\\
&&~~~~~~~~~~~~~~~~~~~~~~~~~~~~~~~~~~~~~~
=\sum_{i=1}^{n}p_{\mu}^{(i)}
|\vec{p}^{~(1)},\ldots,\vec{p}^{~(k)},\ldots,\vec{p}^{~(n)}>\;.\label{pstate}
\end{eqnarray}
We also get from (\ref{stanccr}) the known bosonic symmetry property
\begin{equation}
|\vec{p}^{~(1)},\ldots,\vec{p}^{~(i)},\ldots,\vec{p}^{~(j)},\ldots,\vec{p}^{~(n)}>
=
|\vec{p}^{~(1)},\ldots,\vec{p}^{~(j)},\ldots,\vec{p}^{~(i)},\ldots,\vec{p}^{~(n)}>\;.
\label{pstate}
\end{equation}

In order to complete the structure of $\kappa$-deformed Fock space
we should define dual vectors and scalar product. We define the dual
space in analogy to the relations (\ref{state})
\begin{equation}
<\vec{k}^{(1)},\ldots,\vec{k}^{(n)}| = <0|a_\kappa^\dag
(k_0^{(1)},\vec{k}^{(1)})\circ \ldots \circ a_\kappa^\dag
(k_0^{(n)},\vec{k}^{(n)})\;. \label{dual}
\end{equation}
We define $\kappa$-deformed scalar product of the basic vectors
(\ref{state}) and (\ref{dual}) as follows
\begin{eqnarray}
&&<\vec{k}^{(1)},\ldots,\vec{k}^{(m)}|\vec{p}^{~(1)},\ldots,\vec{p}^{~(n)}>_{\kappa}
:=\; <\vec{k}^{(1)},\ldots,\vec{k}^{(m)}|\circ
|\vec{p}^{~(1)},\ldots,\vec{p}^{~(n)}> =~~~~~~~~~~~~~~~~~~~~~~~~\nonumber\\
&&~~~~~~~~\nonumber \\
&&~~~~~ =<0|a_\kappa^\dag (k_0^{(1)},\vec{k}^{(1)})\circ \ldots
\circ a_\kappa^\dag (k_0^{(m)},\vec{k}^{(m)})\circ
a_\kappa(p_0^{(1)},\vec{p}^{~(1)})\circ \ldots \circ
a_\kappa(p_0^{(n)},\vec{p}^{~(n)})|0>\;, \label{scalar}
\end{eqnarray}
and, using (\ref{stanccr}), it is easy to show that
\begin{eqnarray}
&&<\vec{k}^{(1)},\ldots,\vec{k}^{(m)}|\vec{p}^{~(1)},\ldots,\vec{p}^{~(n)}>_{\kappa}
= ~~~~~~~~~~~~~~~~~~~~~~~~~~~~~~~~~~~~~~~~~~~~~~~~~~~~~~~~\nonumber\\
&&~~~~~~~~~~~~~~~~~~~~~~~~~~~~~=\delta_{nm}\sum_{{\rm
perm}(i_1,\ldots,i_n)}
\delta^{(3)}(\vec{p}^{~(1)}-\vec{k}^{(i_1)})\cdots
\delta^{(3)}(\vec{p}^{~(n)}-\vec{k}^{(i_n)})\;. \label{scalar11}
\end{eqnarray}
We see therefore that the fourmomentum eigenvalues and scalar
products in $\kappa$-deformed Fock space are identical with the ones
characterizing standard bosonic Fock space. We see also from
(\ref{pstate}) that the statistics of n-particle state remains
bosonic\footnote{We define the statistics as define by the symmetry
properties of n-particle states.}.

It appears that the notion of $\kappa$-multiplication is very
useful. It describes the $\kappa$-deformation of the oscillators
algebra as well as the operator-valued metric defining scalar
product in $\kappa$-Fock space.

\section{$\kappa$-deformed interaction vertex in $\lambda\varphi^4$ theory.}

In order to define the perturbative vertex in noncommutative
$\lambda\varphi^4$ theory we shall integrate over space-time the
following product of $\kappa$-deformed free fields
\begin{eqnarray}
&&\varphi(x)\star_{\kappa}\varphi(x)\star_{\kappa}\varphi(x)\star_{\kappa}\varphi(x)
=\int\;{\prod\limits_{k=1}^{4}} d^4p^{(k)}\; \delta
\left(C^2_{\kappa} (\vec{\cal{P}}_4^{(k)},p_0^{(k)}) - M^2\right) \;
A(p_0^{(k)},\vec{p}^{~(k)})\cdot\;~\nonumber \\
&&~~~~~~~~~~~~~~~~~~~~~~~~~~~~~~~~~~~~~~~~~~~~~~~~~~~~~~~~~~~~~~~~~~\cdot{\rm
e}^{ip_\mu^{(1)}{x}^\mu}\star \cdots \star {\rm
e}^{ip_\mu^{(4)}{x}^\mu} = \nonumber\\
&&~~~~~~~~~~~~~=\int{\prod\limits_{k=1}^{4}}d^4p^{(k)}\; \delta
\left(C^2_{\kappa} (\vec{{p}}^{~(k)},p_0^{(k)}) - M^2\right)
\;A(p_0^{(1)},\vec{p}^{~(1)})\circ \cdots \circ
A(p_0^{(4)},\vec{p}^{~(4)})\cdot\; \nonumber \\
&&~~~~~~~~~~~~~~~~~~~~~~~~~~~~~~~~~~~~~~~~~~~~~~~~~~~~~~~\cdot{\rm
exp}\left({i{\sum\limits_{k=1}^{4}}p_{\mu}^{(k)}{x}^\mu}\right)\;.\label{vertex1}
\end{eqnarray}
Performing the integration, one gets $(\textsc{p}_\mu =
{\sum\limits_{k=1}^{4}}p_{\mu}^{(k)})$
\begin{eqnarray}
&&\int\;d^4{x}\;\varphi(x)\star_{\kappa}\varphi(x)\star_{\kappa}\varphi(x)\star_{\kappa}\varphi(x)
= \int\;d^4{x}\;\int\;{\prod\limits_{k=1}^{4}}\;d^4p^{(k)}\delta
\left(C^2_{\kappa} (\vec{p}^{~(k)},p_0^{(k)}) - M^2\right) \;{\rm
e}^{i \textsc{p} x}\cdot \;\;\;\;\;\nonumber\\
&&\hspace{6cm}\cdot\;A(p_0^{(1)},\vec{p}^{~(1)})\circ \cdots \circ
A(p_0^{(4)},\vec{p}^{~(4)}) = \label{vertex2}\\
&&=\int\;{\prod\limits_{k=1}^{4}}\;\;d^4p^{(k)}\;\delta
\left(C^2_{\kappa} (\vec{p}^{~(k)},p_0^{(k)}) -
M^2\right)\;\delta^{(4)}(\textsc{p}_{\mu})\;
\cdot\;A(p_0^{(1)},\vec{p}^{~(1)})\circ \cdots \circ
A(p_0^{(4)},\vec{p}^{~(4)})\;.\nonumber
\end{eqnarray}

From (\ref{vertex2}) we see that the fourmomentum Dirac delta is
classical and describes Abelian conservation law of fourmomenta at
the vertex.

The momentum space formula (\ref{vertex2}) is the proper input
describing $\kappa$-deformed Feynman diagrams in $\lambda\varphi^4$
theory. Because the $\kappa$-deformed mass-shell (\ref{casimir}) due
to the formula \cite{{nlit3}}\footnote{We apply the formula
(\ref{sh}) by putting $x=\frac{p_0}{2\kappa}$.}
\begin{equation}
\frac{\sinh x}{x} = {\prod\limits_{n=1}^{\infty}} \left( 1+
\frac{x^2}{n^2\pi^2} \right) = {\prod\limits_{n=1}^{\infty}} \left(
1+ \frac{ix}{n\pi} \right)\left( 1- \frac{ix}{n\pi} \right)\;,
\label{sh}
\end{equation}
contains infinite number of complex-conjugated poles (see e.g.
\cite{{lit16}}, \cite{{nlit4}}), the problem of defining the
$\kappa$-deformed propagator requires special care, in particular,
the appropriate notion of time ordering. Indeed, let us observe that
(see e.g. \cite{{nlit5}})
\begin{equation}
2\kappa~{\sinh\left(\frac{P_0}{2\kappa}\right)}~f(t) =
\frac{f(t+\Delta t) - f(t - \Delta t)}{\Delta t}~~~~;~~~~\Delta t =
\frac{i}{2\kappa}\;, \label{diff}
\end{equation}
if $P_0 = \frac{1}{i}\partial_t$. We see that in $\kappa$-deformed
field theory the continuous time derivatives are replaced by finite
difference equations and the $\kappa$-deformed Green functions (see
e.g. (\ref{delta}) for equal time limit) satisfy also the finite
time difference equations.

The problem of space-time picture of $\kappa$-causality and proper
definition of $\kappa$-deformed Feynman propagator is now under
consideration.

\section{Outlook.}

Let us recall our main result. We have introduced new
$\kappa$-deformation of creation and annihilation operators, which
inserted in the free quantized fields provide deformed fields with
c-number $\kappa$-deformed Pauli-Jordan commutator function
(\ref{micro1}). It appears that $\kappa$-deformation of free fields
is equivalently described by a new star product $\star_{\kappa}$
(see (\ref{newstar}) and (\ref{newstar1})). The $\kappa$-deformed
oscillators were used for the construction of $\kappa$-deformed Fock
space with multiparticle states, described by momenta which are
added by using the Abelian addition law.

We would like to point out that due to the modification of
$\kappa$-deformed mass-shell deltas (see (\ref{newstar1})) the
$\kappa$-deformed oscillators entering the algebra in Sect. 4 are
beyond the standard $\kappa$-deformed mass-shell. Our modification
of mass-shell condition is however essential in the derivation of
c-number field commutator (\ref{microcausality1}). We recall that
recently (see \cite{ArMar}, \cite{3337}, \cite{3338}) there were
proposed analogous forms of binary algebras for $\kappa$-deformed
oscillators, where  similarly as in our case,  there is  assumed the
modification of three-momentum dependence under the exchange of two
creation (annihilation) operators. It should be stressed however that in
\cite{ArMar}, \cite{3337} the fourmomenta describing the arguments
of all $\kappa$-deformed oscillators in binary algebraic relations
are put on-shell. It can be argued for more general $\kappa$-deformed statistics \cite{ostatnia} that
 specific modification of mass-shell condition for the
$\kappa$-oscillators in binary relations  is necessary for the derivation of the 
c-number commutator for the free $\kappa$-deformed fields.

Our aim is to obtain the perturbative framework for the
$\kappa$-deformed local field theory. We conjecture that the Feynman
rules differs from the standard field framework only by different
choice of the propagators, which satisfy the inhomogeneous
$\kappa$-deformed Klein-Gordon equation. For this purpose one has to
derive in the commutative framework with $\star_{\kappa}$
multiplication the Gell-Mann-Low expansion for $\kappa$-deformed
Green functions and the $\kappa$-deformed Dyson formula. For the
applications it is important to extend the framework to fermionic
Dirac and vectorial gauge fields. It should be recalled that the
$\kappa$-deformed free Dirac and Maxwell fields were already
considered before \cite{klm}. These problems are now under
consideration.

\section*{Acknowledgements}
Two of the authors (M.D. and J.L.) would like to thank P.P. Kulish
for valuable comments.

\end{document}